\pdfoutput=1
\documentclass[10pt, conference]{IEEEtran}

\usepackage{cite}

\ifCLASSINFOpdf
   \usepackage[pdftex]{graphicx}
   \usepackage{epstopdf}
  \graphicspath{{../pdf/}{../jpeg/}}
  \DeclareGraphicsExtensions{.pdf,.jpeg,.png,.eps}
\else
   \usepackage[dvips]{graphicx}
  \graphicspath{{../eps/}}
  \DeclareGraphicsExtensions{.eps}
\fi

\usepackage{color,soul}

\usepackage{amsmath}
\usepackage{amssymb}
\usepackage{amsthm}
\usepackage{algorithm}
\usepackage{algpseudocode}

\ifCLASSOPTIONcompsoc
  \usepackage[caption=false,font=normalsize,labelfont=sf,textfont=sf]{subfig}
\else
  \usepackage[caption=false,font=footnotesize]{subfig}
\fi

\usepackage{balance}

\usepackage{cite}
\usepackage{amsmath,amssymb,amsfonts}
\usepackage{algorithm}
\usepackage{algpseudocode}
\usepackage{graphicx}
\usepackage{textcomp}
\usepackage{xcolor}
\usepackage{verbatim}
\usepackage[utf8]{inputenc}
\usepackage{bm}

\newtheorem{theorem}{Theorem}

\newtheorem{definition}[theorem]{Definition}
\usepackage[T1]{fontenc}

\begin{document}


\title{Entanglement Routing over Quantum Networks Using Greenberger-Horne-Zeilinger Measurements }

\author{\IEEEauthorblockN{
Yiming Zeng\IEEEauthorrefmark{3},
		Jiarui Zhang\IEEEauthorrefmark{3},
		Ji Liu,
		Zhenhua Liu,
		Yuanyuan Yang
}
	\IEEEauthorblockA{
		Stony Brook University, Stony Brook, NY 11794, USA}
	\{yiming.zeng, jiarui.zhang.2, ji.liu, zhenhua.liu, yuanyuan.yang\}@stonybrook.edu
}

\IEEEoverridecommandlockouts
\IEEEpubid{\makebox[\columnwidth]{
\IEEEauthorrefmark{3} Both authors contributed equally to this research. \hfill
} \hspace{\columnsep}\makebox[\columnwidth]{ }}

\maketitle

\begin{abstract}
Generating a long-distance quantum entanglement is one of the most essential functions of a quantum network to
support quantum communication and computing applications.
The successful entanglement rate during a probabilistic entanglement process decreases dramatically with distance, and swapping is a widely-applied quantum technique to address this issue. 
Most existing entanglement routing protocols use a classic entanglement-swapping method based on Bell State measurements that can only fuse two successful entanglement links. This paper appeals to a more general and efficient swapping method, namely $n$-fusion based on Greenberger-Horne-Zeilinger measurements that can fuse $n$ successful entanglement links, to maximize the entanglement rate for multiple quantum-user pairs over a quantum network. We propose efficient entanglement routing algorithms that utilize the properties of $n$-fusion for quantum networks with general topologies. Evaluation results highlight that
our proposed algorithm under $n$-fusion can greatly improve the network performance compared with existing ones.

\end{abstract}


\section{Introduction}
Quantum information science is viewed as the next scientific breakthrough that will propel scientific and
economic developments for the whole society in the near future, since quantum applications have shown capabilities far beyond the traditional approaches. 
Specifically, quantum computing algorithms can lead to an exponential speedup compared to traditional computing algorithms; see for example Shor's algorithm~\cite{lanyon2007experimental} and quantum linear system algorithms~\cite{harrow2009quantum}. Quantum computing algorithms also allow information to be generated, stored, and transmitted at a high level of privacy, security, and computational clout that is impossible to achieve with today’s traditional methods~\cite{gisin2007quantum}. 
The great potential of quantum applications has attracted considerable research attention in academia and huge investments from governments and industries.
For instance, the US Department of Energy has announced that it would provide 625 million US dollars over the next five years to support multidisciplinary quantum information science \cite{DOE}. The European Commission has launched a ten-year, one-billion-euro, flagship project to boost European quantum technology research~\cite{FUND1}. 

In the broad context of quantum information science, quantum networks are expected to be promising next-generation networks~\cite{wang2022application}. 
Several trail quantum networks have been deployed in labs, such as long-distance link (40 kilometers) teleportation over the fiber link~\cite{valivarthi2020teleportation}, the mobile quantum network~\cite{liu2021optical}, and the integrated entanglement system through satellites which can support the entanglement over 4600 kilometers~\cite{chen2021integrated}. 
A fundamental feature of quantum networking, termed entanglement, is important for both formal analysis and physical implementation of quantum computing and communication. 
"Entanglement" refers to a group of qubits (a basic unit to represent quantum information) whose quantum states such as position, momentum, spin, polarization, etc. are correlated, even when the particles are separated by a large distance.
Entanglement between qubits is an essential component of most quantum applications in quantum networks. 

Supporting long-distance entanglement between qubits is critical for quantum networks. 
However, the successful entanglement rate decreases dramatically as the distance between qubits increases. 
Meanwhile, processors in the same group trying to be entangled may be too distant from each other to be directly connected through links. 
{\em Entanglement-swapping} is an important method that can establish an entanglement path between those pairs of quantum processors that had not shared an entanglement. 
In this way, certain processors in a network that do not participate in the current entanglement can work as relays through entanglement-swapping to supply end-to-end entanglement. 
Such a processor is called a switch. This paper studies a key problem in quantum networks, called  {\em entanglement routing}, whose goal is to achieve efficient long-distance entanglement over a quantum network through entanglement-swapping. 

\begin{figure}[ht] 
    \vspace{-.2in}
    \centering
    \subfloat[]{
       \includegraphics[width=0.3\columnwidth]{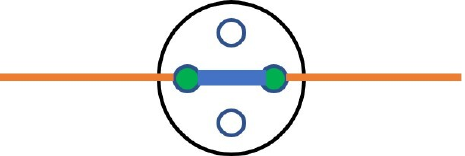} \label{pic: bsm}
    }
    \hspace{.4in}
    \subfloat[]{
        \includegraphics[width=0.3\columnwidth]{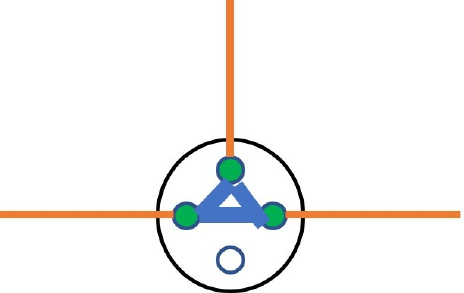}\label{pic:ghz-1}
    }
    \hspace{-.5in}
    \caption{(a) A BSM measurement in the switch that fuses two quantum links by connecting two qubits.  (b) A 3-GHZ measurement in a switch that fuses three quantum links by connecting three qubits. In both figures, the small blank circle in the switch denotes free qubits that are not entanglement, the small green circle in the switch denotes entangled qubits,
    the orange line indicates the quantum links to be fused, and the blue line shows the connection between qubits to fuse quantum links.  }

\end{figure}

The entanglement routing problem has recently drawn great attention and a few motivating results have been obtained.
Studies have proposed entanglement routing algorithms and corresponding theoretical analyses for some special classes of network topologies \cite{pant2019routing,li2021effective,chakraborty2019distributed,vardoyan2019stochastic,shchukin2019waiting,das2018robust}. Improvements for arbitrary network topologies have been made in \cite{zeng2022multi,shi2020concurrent,zhang2021fragFmentation}. 
In these existing quantum routing algorithms, however, switches are restricted to perform a classic swapping method~\cite{pan1998experimental} that implements so-called Bell State Measurements (BSMs). A BSM is based on two qubits and thus can only fuse two entanglement links. 
Figure~\ref{pic: bsm} shows an example that BSM can fuse two quantum links by entangling two qubits in the processor.

A few recent works~\cite{patil2022entanglement,patil2021distance} allow switches to perform multi-qubit joint Greenberger-Horne-Zeilinger(GHZ) measurements in the GHZ basis,
which can fuse $n \ge 2$ successful entanglement links simultaneously and thus called $n$-fusion. Fig.~\ref{pic:ghz-1} illustrates a case where a processor takes 3-GHZ measurements that can enable the processor to fuse three successful entanglement links. Clearly, 
$n$-fusion is a more general entanglement-swapping method subsuming the classic BSM-based swapping method (i.e., $2$-fusion) as a special case.  
Implementing $n$-fusion with $n \geq 3$ is in principle no harder than 2-fusion (BSMs) in switch memories~\cite{patil2021distance}. 
Meanwhile, $n$-fusion provides the network with more choices and flexibility for individual switches to generate entanglement for quantum-users, which may greatly impact the network entanglement performance.   
However, the existing works~\cite{patil2022entanglement,patil2021distance} only consider a single quantum-user pair in special grid networks, and their results are limited to 3- and 4-fusion measurements.
Therefore, a comprehensive study for quantum entanglement routing based on $n$-fusion is still missing in the literature. 

To fill the gap just described, this paper considers general network topologies, $n$-fusion entanglement-swapping with arbitrary integer $n\ge 2$, and multiple quantum-user pairs simultaneously.
To the best of our knowledge, this is the first work of a comprehensive entanglement routing design over a general quantum network topology with multiple quantum-user pairs where switches can take a general entanglement-swapping method, $n$-fusion, with novel models, metrics, and algorithms. 
The specific goal is to maximize the entanglement rate of the network, i.e., the expected number of shared quantum states between quantum-user pairs. Our contributions are as follows:
\begin{enumerate}
    \item We present a novel comprehensive entanglement routing model and detailed entanglement process where switches take $n$-fusion for multiple quantum-user pairs to share quantum states in a general network topology. 
    \item We propose novel routing metrics under $n$-fusion measurement to evaluate network performance and optimize them by algorithm design.  
    \item We design efficient entanglement routing algorithms to allocate paths for sharing quantum states between quantum-user pairs. 
    \item We perform a series of simulations that demonstrate that under the same network setting, our novel $n$-fusion algorithm not only improves the performance up to $6\times$ compared with the existing classic swapping algorithms but also significantly outperforms
    the existing $n$-fusion ones in terms of the network entanglement rate. 

\end{enumerate}
This paper provides a foundation for further investigation of $n$-fusion as a swapping method to support quantum applications. 

The rest of this paper is organized as follows. Section~\ref{sec: background} introduces the necessary quantum background needed in the paper and Section~\ref{sec: network model} presents the network model. 
We propose algorithm designs in Section~\ref{sec: algorithms}. 
We conduct extensive simulations to discuss and analyze the performance of our proposed algorithms  and compare them with previous ones in Section~\ref{sec: simulation}, followed by related work in Section~\ref{sec: related work} 
the conclusion in Section~\ref{sec: conclusion}.

\section{Background}\label{sec: background}
In this section, we will introduce some important quantum terminologies and mechanisms that we will use in this paper. 

\subsection{Basic Terminologies}
\textbf{Qubit:} A qubit is a basic unit to represent quantum information, which can be an electron or a photon or a nucleus from an atom, and be described by its state~\cite{van2014quantum}. Different from an {\em ebit} in classical computing which represents 0 or 1, a qubit can present a coherent superposition of both 0 and 1. 
\noindent

\textbf{Entanglement:} Entanglement is a phenomenon that a group of qubits expresses a high correlation state which cannot be expressed by the states of individual qubits. $n$ qubits can maximally be entangled as a $n$-GHZ state, i.e., $\frac{|0\rangle^{\otimes n}+|1\rangle^{\otimes n}}{\sqrt{2}}$. The Bell State that contains the exact two qubits can be viewed as a special case of the $n$-GHZ state, where $n=2$.  

\subsection{Entanglement-swapping}
\begin{figure}[htbp]
\centerline{\includegraphics[width=0.3\textwidth]{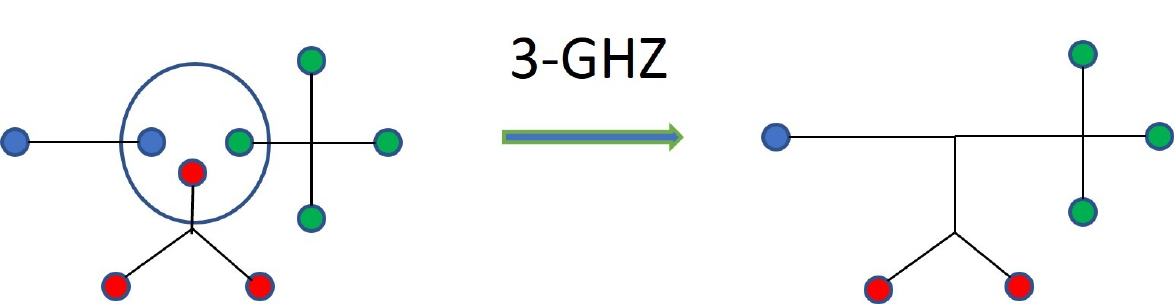}}
\caption{A 3-fusion swapping in a processor over 3 independent processor sets.}
\label{pic: GHZ1}
\end{figure}

Entanglement-swapping is a quantum operation in which if two processors each has a different qubit entangled with another common processor, then the qubits of these two processors can be entangled directly with the help of the common processor. 
In this paper, we consider a general entanglement-swapping operation called $n$-fusion~\cite{patil2022entanglement,patil2021distance}. 
$n$-fusion can project $n$ measured qubits onto one of the $2^n$ GHZ states by taking GHZ measurements. 
In a quantum network, this operation allows switches to fuse $n$ successful entanglement links simultaneously.  
Figure~\ref{pic: GHZ1} shows an example that a three-party processor set gets entangled as a 6-GHZ state through GHZ measurements in a switch. 
For the interior of a switch, $n$-fusion entangles its qubits as shown in Figure~\ref{pic:ghz-1}. 

When $n=1$, the operation takes as a single-qubit Pauli measurement~\cite{crawford2021efficient}, which removes a qubit in the measurement axis and results in a $(n-1)$-GHZ state from a $n$-GHZ state. 
When $n=2$, the operation is a classic swapping procedure by taking BSMs over two qubits that were produced independently and never previously interacted~\cite{pan1998experimental}. 
Taking a $n \geq 3$-fusion is in principle no harder than a classic swapping (i.e., $n=2$)~\cite{patil2021distance}, but it will bring more choices and flexibility for the entanglement among switches. 

The entangled quantum state shared between a quantum-user pair can support teleportation which is a quantum operation to secretly transfer a quantum state from one processor to another processor. The transmitted quantum information can be either a single qubit~\cite{riebe2004deterministic} or a GHZ state~\cite{saha2012n}.

\section{Quantum Network Model} \label{sec: network model}
In this section, we will introduce the quantum network model based on facts from practical physical experiments, and related studies about quantum entanglement routing to reflect practical quantum networks. 



\subsection{Network Component} 
We consider a general network topology where switches take $n$-fusion entanglement-swapping for the entanglement. 
We first introduce four main components of the quantum network. 

\textbf{1. Quantum-users.} 
A quantum-user is a quantum processor that tries to establish entanglement with other processors by sharing the same quantum state. 
The processor includes two sets of qubits: one set for performing quantum computing tasks and the other for communication, specifically for entangling with other quantum-users through swapping. 
In this paper, we consider a scenario where a quantum state can only be shared between two quantum-users, enabling them to perform teleportation.
There are multiple quantum-user pairs in the network. 
Each quantum-user can use different qubits to entangle with different quantum-users simultaneously.  
However, each qubit of a quantum-user can only at most be involved in one quantum state. 
The set of quantum-users is denoted as $\mathcal{U}=\{u_i\}_{i=1}^U$. 

\textbf{2. Quantum switches.}
Quantum switches are quantum processors that work as relays to support remote entanglement through entanglement-swapping. 
Unlike quantum-users, which have both computing and communication qubits, quantum switches only have qubits for communication and are used solely for entanglement-swapping. 
The set of switches is denoted as $\mathcal{V}=\{v_i\}_{i=1}^V$.
In this model, we assume that switches are capable of performing up to $c_v$-qubit GHZ measurement, where $c_v$ is the number of qubits in the solid memory of switch $v \in \mathcal{V}$. Additionally, the switches are equipped with traditional computing and communication devices, allowing them to communicate with one another via optical fibers and perform traditional computing tasks. 

\textbf{3. Quantum links.} 
The quantum link over the optical fiber connects two switches to support the entanglement. 
If the quantum link is successfully established, the neighboring switches share a Bell pair $\frac{|00\rangle+|11\rangle}{\sqrt{2}}$.  
The successful entanglement probability over the quantum link is determined by the link length and the physical material of the optical fiber, represented by $p=e^{-\alpha L}$, where $\alpha$ is a constant dependent on the physical material and $L$ is the length the quantum link between switches.  
The network is abstracted as an undirected graph, denoted as  $G=(\mathcal{\overline{V}}, \mathcal{E})$, where $\mathcal{\overline{V}}=\{\mathcal{U} \cup \mathcal{V}\}$ denotes the set of quantum switches and quantum users. 
Let $e_{ij}$ denote the edge between switches or users $v_i \in \mathcal{\overline{V}}$ and $v_j \in \mathcal{\overline{V}}$, and $\mathcal{E}= \{ e_{ij}\}\subset \{(v_i,v_j)\;: \; v_i, v_j \in \mathcal{\overline{V}} \}$ denote the set of edges. 
Since several optical fiber cables can be placed between switches and each cable has independent cores that can be used as quantum links, there could be multiple quantum links over the same edge for one quantum state. These quantum links for one quantum state are referred to as a quantum channel. 
Additionally, multiple quantum channels for different quantum states can coexist on the same edge.  
The optical fibers can also be used to transmit traditional information over the network, i.e., an e-bit that presents 0 or 1.

\textbf{4. A center server for traditional computing and communication}
The center server plays a crucial role in maintaining network information such as network topology, connections, etc., and relaying information to the switches. The communication process between the center server and the switches may operate in an honest-but-curious manner, meaning that the server will not have access to the content of the communication.
The center server will also perform computational tasks to support the entanglement process, such as pre-calculating routes for quantum-user pairs.

\begin{figure}[htbp]
\vspace{-1.2in}
\centerline{\includegraphics[width=0.45\textwidth]{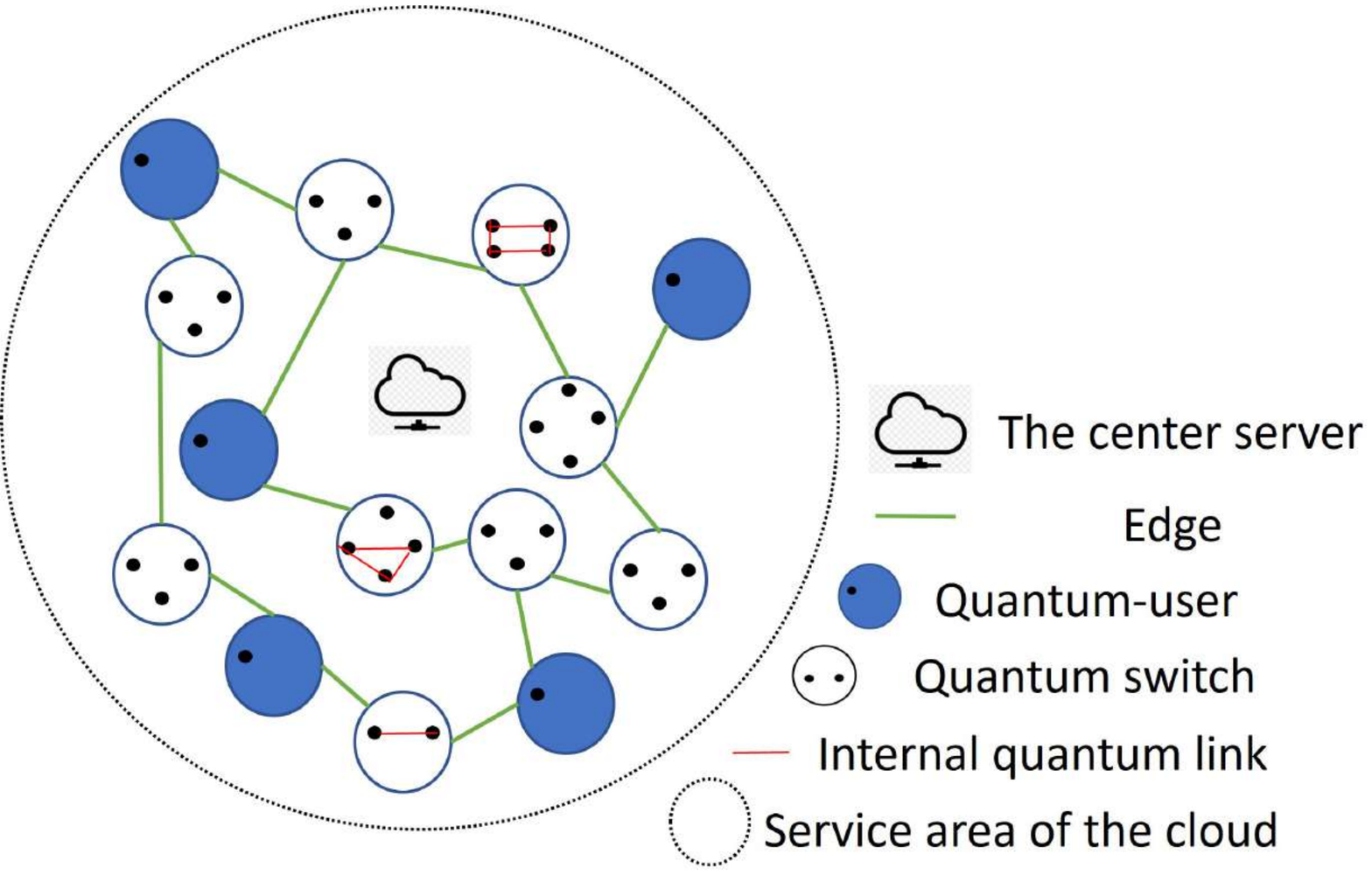}}
\vspace{-1.3in}
\caption{An Example of Network.}
\label{fig:network}
\vspace{-0.1in}
\end{figure}

\textbf{Network topology.}
We model a quantum network with $U$ quantum-users and $V$ switches
as an undirected graph $G=(\mathcal{\overline{V}},\mathcal{E})$, where $\mathcal{\overline{V}}=\{\mathcal{U} \cup \mathcal{V}\}$, and $\mathcal{E}= \{ e_{ij}\}\subset \{(v_i,v_j)\;: \; v_i, v_j \in \mathcal{\overline{V}} \}$ denotes the set of the quantum links.  
The topology of the graph is arbitrary. 
Figure~\ref{fig:network} shows an example of the proposed quantum network. 
The topology of the quantum network is relatively stable, with network information being readily available to all switches.

\subsection{Entanglement Process} \label{subsec: entanglement process}
In this model, we briefly outline the entanglement process in which switches utilize $n$-fusion entanglement-swapping. 


Phase I is to design detailed routes for the entanglement. The routes are computed by classic computing devices in the center server since the computing time for tasks such as routing problems is currently still much shorter in the classical computing domain than in the quantum domain. 
The center server uses the available network information, including quantum states to be shared for quantum-user pairs, network topology (switch placement and connections), and switch information (the number of qubits in each processor), to design the routes.  

Phase II is the preparation phase for the entanglement. The network sends designed routes to switches. 
All switches will be time-synchronized to ensure that all links are available simultaneously for the selected paths. 

Phase III is to take entanglement. First, the network attempts to generate quantum entanglement over the quantum links based on the fixed routing paths calculated in Phase I. Second, switches take entanglement-swapping operations to fuse successful entanglement links for the entanglement.

\subsection{Routing Matrices}
We define the \emph{entanglement rate} as the expected number of quantum states (including Bell states and GHZ states) shared between quantum-user pairs in the quantum network under $n$-fusion to quantify the performance. 

\begin{figure}[htbp]
\vspace{-0.5in}
\centerline{\includegraphics[width=0.25\textwidth]{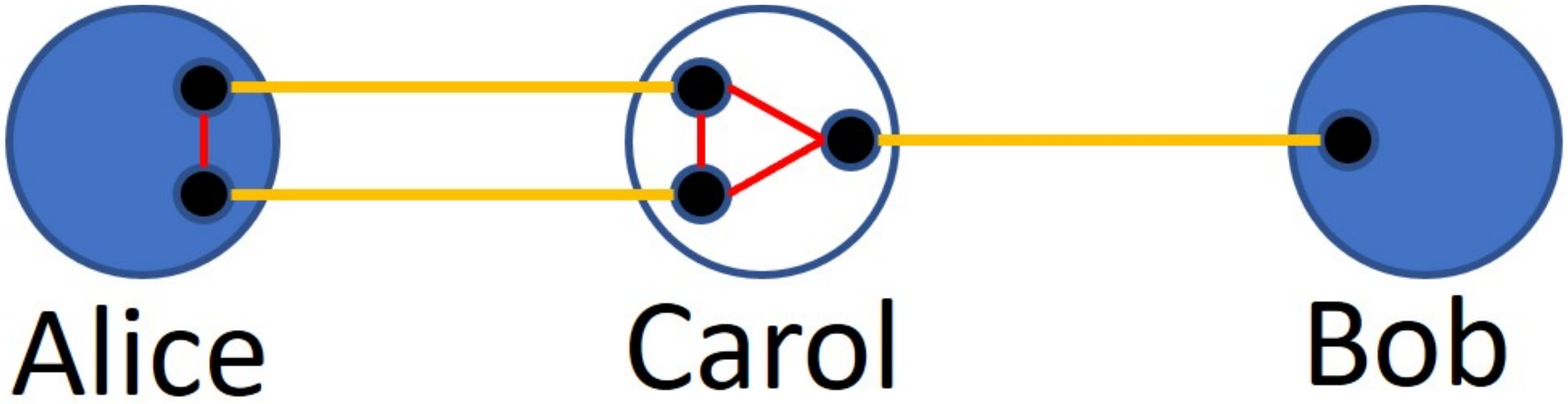}}
\vspace{-0.5in}
\caption{Entanglement example 1: a path. Qubits in black are entangled through red internal links in switches and orange links between switches.}
\label{pic: matrics1}
\vspace{-0.1in}
\end{figure}

\textbf{{\em Entanglement rate} of a quantum channel.} 
Given two neighboring switches $v_i, v_j$, let $w$ denote the width of the quantum channel which indicates the number of parallel quantum links placed between two switches for one quantum state to be shared. 
As shown in Figure~\ref{pic: matrics1}, the width of the edge between Alice and Carol is 2.  
The entanglement rate  between two adjacent switches equals the probability to build at least one successful entanglement link for a quantum state, i.e., $P=1-(1-p_{ij})^{w}$, where $p_{ij}=e^{-\alpha L_{ij}}$ is the successful entanglement probability over a single link with the Euclidean length $L_{ij}$.  
In Figure~\ref{pic: matrics1}, let $p$ denote the successful entanglement probability of a single quantum link and all links have the same rate, then, the entanglement rate between Alice and Carol is $1-(1-p)^2$. 

\textbf{{\em Entanglement rate} of a path.}
For a path $A=\{v_1, v_2, \cdots, v_l\}$, where $l$ denotes the distance of $A$ (i.e., the number of edges), the entanglement rate for a quantum state is determined by the successful entanglement-swapping probability of each switch in a path, and the entanglement rate of each edge. 
Formally, the entanglement rate of path $A$ is $P_A=\Pi_{i=2}^{l-1}q_i \Pi_{j=1}^{l}P_{j(j+1)}$.  
In Figure~\ref{pic: matrics1}, the entanglement rate of a path between Alice and Bob is $(1-(1-p)^2)pq$, where $q$ is the successful entanglement-swapping probability of Carol to take $n$-fusion to fuse three quantum links. 

\textbf{{\em Entanglement rate} of a flow-like graph.} 
To generate a quantum state shared between a quantum-user pair, a flow-like graph may be constructed, which is defined as follows:
\begin{definition}
A flow-like graph is a symmetric flow graph connecting two quantum-users for one shared quantum state, which contains paths sharing overlapped nodes or edges. In a symmetric flow graph, the overlapped node contained by more than one path for the same shared quantum state is called the branch node. 
\end{definition}

\begin{figure}[htbp]
\vspace{-0.1in}
\centerline{\includegraphics[width=0.35\textwidth]{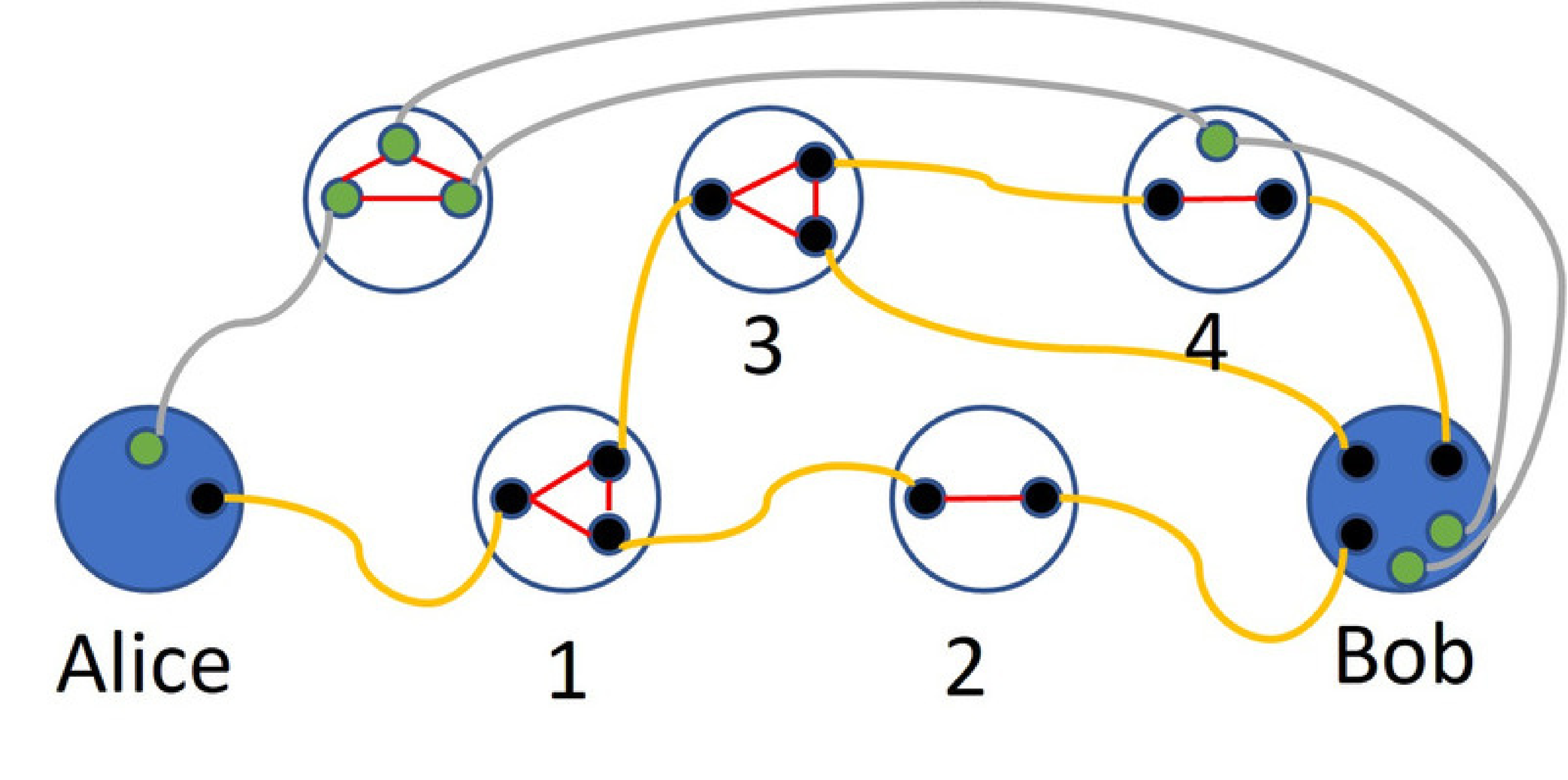}}
\caption{Entanglement example 2: two flow-like graphs for two shared quantum states between a quantum-user pair. qubits in green are entangled through red internal links in switches and gray links between switches.}
\label{pic: matrics2}
\end{figure}

An example of flow-like graphs is shown in Figure~\ref{pic: matrics2}.
A switch as a branch node belongs to different paths simultaneously that connect two quantum-users for one quantum state, where each path leads to an independent direction for the entanglement. 
For example, switch 3 is a branch node that has two paths for one shared state whose qubits are in black; but switch 4 is not a branch node even if it belongs to two paths because two paths belong to two different shared quantum states. 
For any two different quantum states between the same quantum-user pair, their flow-like graphs do not share any quantum links including internal links within switches and links between switches on the edge. 

Calculating the entanglement rate for a shared quantum state in a flow-like graph requires traversing every branch node and every path through  every branch node.  
The calculation process starts with one quantum-user traversing all edges and switches one by one. 
Let $\hat{G}=\{U_1, U_2, \varrho\}$ denote a flow-like graph between quantum-user  $U_1$ and $U_2$ for sharing the quantum state $\varrho$. 
We assume that user $U_1$ is the root node. 
Starting from user $U_1$, neighboring nodes of user $U_1$ along the
direction from user $U_1$ to user $U_2$ are called child nodes. 
Every child node of node $A$ has its own child nodes. 
$P_{\{a, b, \varrho\}}$ represents the entanglement rate from node $a$ to node $b$ for $\varrho$. 
When we fix quantum state $\varrho$, the notation is shortened as $P_{a, b}$, where node $a$ and node $b$ are two nodes in $\hat{G}=\{U_1, U_2, \varrho\}$ and node $a$ is has a shorter distance to user $U_1$ compared with node $b$. 
$\mathcal{F}_{a,b}$ denotes the set of all paths starting from node $a$ and ending at node $b$ in $\hat{G}=\{U_1, U_2, \varrho\}$. 
$\mathcal{C}_a$ denotes the set of all child nodes of node $a$ in a symmetric flow graph. 
Then the entanglement rate of $\hat{G}=\{U_1, U_2, \varrho\}$ can be computed via a recursive process:
\begin{align}\label{eq: entanglement rate}
    P_{\{U_1, U_2\}}=1-\prod_{u \in \mathcal{C}_{U_1}}\big(1-P_{\{U_1, u\}}P_{\{u, U_2\}} \big).
\end{align}

\subsection{The Entanglement Routing Problem}
This work explores an entanglement routing problem in the quantum network where switches can take $n$-fusion entanglement-swapping through GHZ measurements. 
We consider a given quantum network with an arbitrary network graph $G=(\mathcal{\overline{V}}=\{\mathcal{U}\cup \mathcal{V}\}, \mathcal{E})$, where a set of quantum-user pairs $\{\langle u_i, u_j\rangle, u_i, u_j \in \mathcal{U}\}$ tries to form entangled quantum states (including Bell states and GHZ states) with each other. A quantum-user could share different states with different quantum-users simultaneously,  and multiple quantum states could be shared between one quantum-user pair. 
We assume that quantum-users have enough quantum memories (qubits) for the entanglement because a quantum-user can be formed as a virtual quantum machine with a large number of qubits by entangling a group of quantum processors to boost the memory capability. 

Currently, the number of qubits stored in the solid memory of a switch is limited by technical constraints, i.e., up to 127 qubits per switch \cite{chow2021ibm}, and high construction cost, i.e., the average cost to build a single qubit in a quantum processor can be up to 10,000 U.S. dollars~\cite{Jhon2022Million-qubit}. 
However, optical fibers have mature construction technologies with relatively low cost (i.e., 0.5 U.S. dollars per kilometer). 
One optical fiber cable can contain up to 25 cores, and each core can be used as an independent link for the entanglement.
Moreover, multiple optical fiber cables can be placed between quantum switches. 
Therefore, the network has enough edge capacity to satisfy entanglement demands. Let $c_v$ denote the number of qubits in switch $v \in \mathcal{V}$. 
In this paper, we consider the number of qubits in each switch as the main limitation that indicates a switch cannot allocate qubits for the entanglement more than its capacity. Each switch can determine how to allocate qubits for building quantum links. 
We assume that all switches have the same successful entanglement-swapping probability $q \in [0,1]$ for $n$-fusion.
The successful entanglement probability of a quantum link is determined by its Euclidean length, i.e., $P_{ij}=e^{-\alpha L_{ij}}$. 

In this paper, our goal is to maximize the entanglement rate of the network, i.e., the expected number of shared quantum states between quantum-user pairs. The entanglement process is illustrated in Section~\ref{subsec: entanglement process}. 

\section{Entanglement Routing Algorithms}
\label{sec: algorithms}
In this section, 
we will first introduce the challenges to solve the entanglement routing problem and then propose a few main ideas concluded from the observation as principles to design routing algorithms.
Based on the main ideas, we will propose algorithms for entanglement routing. 

\subsection{Challenges}
The key distinction between the problem addressed in this paper and previous studies is that we consider a more general model in which switches can perform $n$-fusion entanglement-swapping.
This method of entanglement-swapping has the following benefits: 
(1) Fusing more quantum links for one quantum state: 
Since $n$-fusion allows for fusing less or equal to $n$ quantum links instead of exactly 2,  the switch has the ability to use qubits more efficiently by allocating more than 2 qubits for a single quantum state that is shared between a quantum-user pair. 
In previous works, a switch can only assign 2 qubits per quantum state. 
(2) Increased flexibility and routing options for entanglement: 
a switch can allocate different qubits to connect with other switches for a single quantum state, providing the network with more flexibility and choices in the entanglement routing design, leading to an improvement in the entanglement rate. 

Meanwhile, two fundamental issues need to be addressed in the entanglement routing algorithm:
\begin{enumerate}
    \item Routing: determining the connections between quantum-user pairs for shared quantum states.
    \item Allocation: assigning qubits in switches for each chosen route.
\end{enumerate}

The challenges for these two issues are two-fold. 

(1) For the first one, it is challenging to design routes for quantum-user pairs under $n$-fusion. The number of possible ways of $n-$fusion routing is far greater than the number of possible ways of classic swapping. 
In the classic swapping, only paths exist between a quantum-user pair, and one path is found for each quantum state. Some existing algorithms like Yen's algorithm~\cite{yen1971finding} and Dijkstra's algorithm~\cite{dijkstra2022note} can be applied directly to find paths for a quantum-user pair.  
However, in this paper, under $n$-fusion, the connection between a quantum-user pair could be a flow-like graph instead of a path. A connection under $n$-fusion can be a combination of multiple paths in classic swapping. 
There could be up to $2^{|\mathcal{V}|!e}$ flow-like graphs and paths between one quantum-user pair in a complete graph (the switches can be selected multiple times), where $|\mathcal{V}|$ is the number of vertices in $\mathcal{G}$, $!$ denotes the factorial, and $e$ is Euler's number\footnote{It is a mathematical constant approximately equal to 2.71828.}. 
Such a huge route set will cause great computational overhead to find high-performance routes under $n$-fusion.

(2) For the second problem, it is hard to determine qubit allocation in switches for routes of different quantum states to maximize the entanglement rate.  
There are two reasons.  
First, the complexity to compute the entanglement rate of routes is high. The expression of the entanglement rate, i.e., Equation~\ref{eq: entanglement rate}, is non-linear in a recursive form. 
Additionally, calculating the entanglement rate of a flow-like graph directly needs to traverse all branches and paths in a recursive way, which will generate significant computational overhead. 
Second, The complexity in designing algorithms for qubit allocation in a flow-like graph is greatly increased by the fact that the number of qubits assigned to a shared quantum state by switches 
may vary at different switches.  
In contrast, in the classic swapping, the number of qubits assigned to a path from different switches is the same. 
As a result, the algorithm should be efficient enough for nodes to design routes among a large of possible ways, and efficiently assign qubits in nodes for complex paths of different pairs.




Overall, the entanglement routing under $n$-fusion is a completely new problem in quantum networking whose complexity is far beyond the case under the classic swapping.


\subsection{Main Ideas} 
Our proposed algorithm is based on the following novel and interesting ideas observed from the unique features of the quantum network. 

\textbf{1. Flow-like graphs versus paths.}
An important advantage of using $n$-fusion over switches is the ability to generate a flow-like graph between a pair of quantum-users for a shared quantum state. 
However, finding flow-like graphs is very challenging, and calculating their entanglement rate has high time complexity. 
To overcome that, we use an alternative method instead of finding a flow-like graph directly. 
A flow-like graph can be viewed as a union set of several paths for the same quantum state. 
Therefore, paths are found first and then merged as a flow-like graph to maximize the entanglement rate. 
The following ideas will mainly focus on paths. 

\textbf{2. `Shorter' paths are preferred.}
Here, we focus on the condition of the 1-width path. This is because a path with a width greater than 1 can be obtained by repeating a 1-width path. 
On the other hand, finding a 1-width path is necessary as a foundation for exploring paths with a width greater than 1.  
The entanglement routing process is probabilistic for both entanglement over quantum links and swapping over switches.
When the width of a path is 1, the entanglement rate of the path is determined by the total length of the path and the number of switches. A path that has a short length but many hops may still have a small entanglement rate. 
Therefore, the term `shorter' path refers to a path with a higher entanglement rate by jointly considering the length and the number of hops. 
A `shorter' path is preferred because it has a larger entanglement rate. 
Moreover, a shorter path will use fewer resources in the network, allowing the network to handle more demands.
%

\textbf{3. Wider paths are preferred.} 
A larger width indicates more quantum links in a quantum channel for a shared quantum state between switches. 
For a quantum channel between two switches in the routing path of a quantum-user pair to share a quantum state, increasing the width can increase the entanglement rate. For example, given a quantum channel with $n$ quantum links, the entanglement rate is $1-(1-p)^n \approx np$. The approximate equality (i.e., $\approx$) holds when $p$ is small, which fits the reality that the successful entanglement probability over optical fiber is typically 0.01\%~\cite{dahlberg2019link}. 

\begin{figure}[ht] 
\vspace{-0.1in}
    \centering
    \subfloat[]{
        \includegraphics[width=0.4\columnwidth]{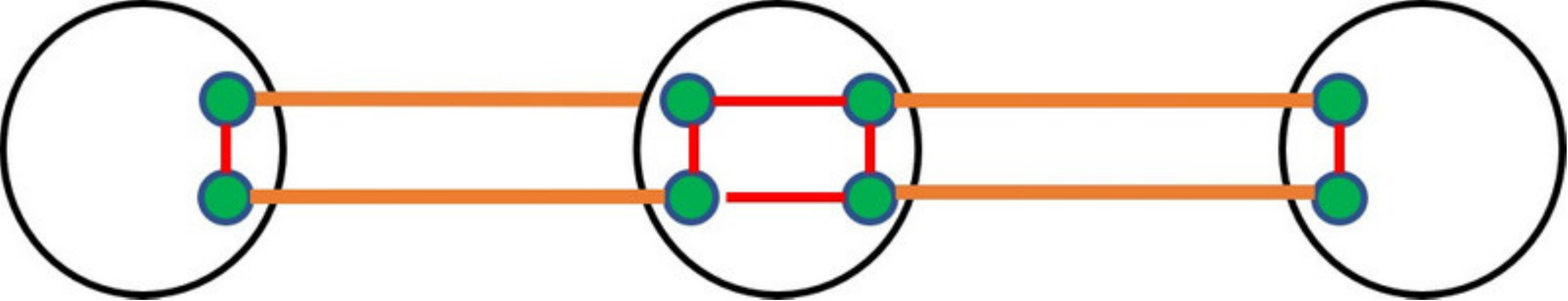} \label{pic: E3}
    }
    \hspace{.1in}
    \subfloat[]{
        \includegraphics[width=0.4\columnwidth]{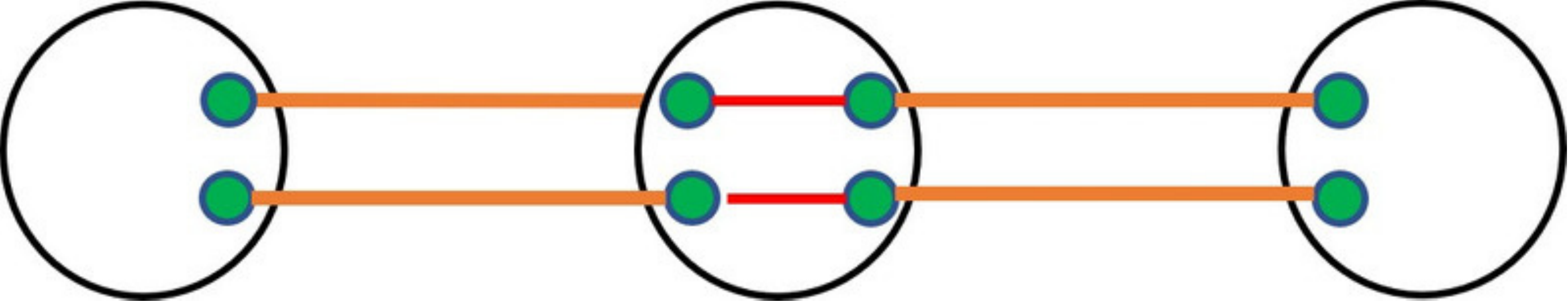}\label{pic: E4}
    }
    \caption{(a) An example of an n-fusion flow graph. (b) An example of two shared states with the classic swapping.}
\vspace{-0.1in}
\label{pic: E34}
\end{figure}

\textbf{4. $n$-fusion is preferred.} 
In a flow-like graph or a path for a quantum-user pair,  using n-fusion to fuse as many quantum links as possible leads to a larger entanglement rate than classic swapping.  
For example, in Figure~\ref{pic: E3}, switches take $4$-fusion for one quantum state whose entanglement rate is $q(1-(1-p)^2)^2$. 
In Figure~\ref{pic: E4}, switches take classic swapping for two quantum states whose entanglement rate is $2p^2q$.
When $p$ is small,  the entanglement rate in Figure~\ref{pic: E4} is less than Figure~\ref{pic: E3}. 
With the same width, length, and hops in a path, $n$-fusion is more possible to have a larger entanglement rate than the classic swapping. 
This conclusion can be extended to a general case for a path with widths $w$ and hops $z$. Under $2w$-fusion, the entanglement rate is $(1-(1-p)^w)^z q^{z-1} \approx (wp)^z q^{z-1}$, where we assume that $p$ is small. 
Under the classic swapping, the entanglement rate is $wp^zq^{z-1}$. The ratio between entanglement rates is $\frac{(wp)^z q^{z-1}}{wp^zq^{z-1}}=w^{z-1}$.
We can conclude that in this graph setting, the entanglement rate under $n$-fusion is larger than the classic swapping.


\subsection{Entanglement Routing Algorithm Design}
\subsubsection{Algorithm Overview}
Based on the main ideas, we present an entanglement routing algorithm as three steps where each step contains one or two sub-algorithms. 

In \textit{Step I}, we will design an algorithm to construct a path set with various widths for different shared quantum states, taking into account network limitations such as the number of qubits in switches.  
Some paths will be selected from this set to be used as routes directly or merged into flow-like graphs for shared quantum states. 

In \textit{Step II}, we will select paths starting from the largest width with the highest entanglement rate one by one and try to merge the selected paths for the same quantum state to save qubit resources in the network. This will result in the construction of flow-like graphs. 

In \textit{Step III}, if there are unused qubits, they will be utilized to form quantum links and added to paths or flow-like graphs designed in \textit{Step II} to increase the entanglement rate.
Each switch will fuse as many quantum links as possible for each shared quantum state according to routes determined by the  entanglement routing algorithm.  



\subsubsection{Step I: Construct a path set}


As previously discussed, flow-like graphs and paths may co-exist between quantum-user pairs for shared quantum states, and it is very challenging to construct flow-like graphs directly. 
To address that, we find potential feasible paths first, and then try to merge paths that belong to the same shared quantum state under the network limitation to maximize the network entanglement rate. 


In Step I, we aim to construct a set of paths with different widths to be used for the entanglement routing. 
The path selection follows the principles of being `shorter' and `wider', which means we will select wide paths with large entanglement rates. 
Since paths in the set are for determining the final entanglement routes, the resources in the network can be repeatedly used in the path selection process. 
To address this, we first propose an algorithm to build a `shortest' path with a fixed width, 
which is to find a path with the largest entanglement rate with a fixed width. 
The details of the algorithm are concluded in Algorithm~\ref{alg-1}. 
Algorithm~\ref{alg-1} is the foundation to select multiple paths with different widths. 
Then, we will combine Algorithm~\ref{alg-1} and Yen's algorithm~\cite{yen1971finding} in Algorithm~\ref{alg-2} to select multiple paths with different widths.

\renewcommand{\algorithmicensure}{\textbf{Input:}}
\renewcommand{\algorithmicrequire}{\textbf{Output:}}
\begin{algorithm}
	\small
	\caption{Largest Entanglement Rate Algorithm}
	\label{alg-1}
	\begin{algorithmic}[1]
		\Ensure $\mathcal{\overline{V}},\mathcal{E}, \langle S, D, \varrho \rangle , \mathcal{Q}, w$
		\Require $A=\{S,a_1,...,a_l,D\},Met_D$
		\State $Queue\xleftarrow{}\{S\}$, $Met_i\xleftarrow{}0 (\forall i\in\mathcal{\overline{V}})$, $Met_S\xleftarrow{}1$
		\If {$Q_S<w$ or $Q_D<w$}
		    \State Return no solutions
		\EndIf
		\While {$Queue\neq \emptyset$}
			\State Select $v_i\in {Queue}$ s.t. $Met_{i}$ is maximum
			\State Remove $v_i$ from $Queue$
			\ForAll {$v_j\in \mathcal{\overline{V}},e_{ij}\in\mathcal{E}$}
				\If {$Met_j < Met_i$ through $e_{ij}$ and $Q_j\geq 2w$}
					\State $Met_j \xleftarrow{} Met_i$ through $e_{ij}$, $pre_j\xleftarrow{}i$
					\If {$v_j\notin Q$}
						\State $Queue\xleftarrow{} Queue\cup v_j$
					\EndIf
				\EndIf
			\EndFor
		\EndWhile
		\State $t\xleftarrow{}pre_D$
		\While {$t\neq S$}
			\State {$a_l\xleftarrow{}v_t,t\xleftarrow{}pre_t,l\xleftarrow{}l-1$}
		\EndWhile
	\end{algorithmic}
\end{algorithm}
\begin{algorithm}
	\small
	\caption{Paths Selection Algorithm}
	\label{alg-2}
	\begin{algorithmic}[1]
		\Ensure $\mathcal{G}=(\mathcal{\overline{V}},\mathcal{E}), \mathcal{S}, \mathcal{D}, \mathcal{M}=\{\varrho_{SD}, S \in \mathcal{S}, D \in \mathcal{D}\}, \mathcal{Q}$, $h$
		\Require $\mathcal{A}_w,\forall w\in[1,\textsc{MAX\_WIDTH}]$
		\ForAll {width $w$ from MAX\_WIDTH to 1}
            \State $Alg1(\mathcal{\overline{V}},\mathcal{E},\langle S, D, \varrho \rangle , \mathcal{Q}, w)$
            \State Obtain path $A$ and metric $Met_D$ by Algorithm 1
			\State $\mathcal{A}_w\xleftarrow{} \emptyset$
            \State $PQueue\xleftarrow{}\{(\emptyset,A,Met_D)\}$
            \While {$PQueue\neq \emptyset$ and $\mathcal{A}_w<h$}
	            \State Select a tuple $(\mathcal{E'},A'=\{a'_1,a'_2,...,a'_l\},Met_D')\in PQueue$ with the maximum $Met_D'$
	            \State $\mathcal{A}_w\xleftarrow{} \mathcal{A}_w \cup A'$
	            \ForAll {$(a'_i,a'_{i+1})\in A'$}
            	    \State $Alg1(\mathcal{\overline{V}},\mathcal{E}-\mathcal{E'}-e,\langle a'_i, D, \varrho \rangle , \mathcal{Q}, w)$
            	    \State Obtain path $A=\{a'_i,a_2,a_3,...,a_l,D\}$ and metric $Met_D$ by Algorithm 1
            	    \State $A\xleftarrow{}\{a'_1,a'_2,...,a'_i,a_2,...,a_l,D\}$
            	    \State $PQueue\xleftarrow{} PQueue\cup (\mathcal{E'}+e,A,Met_D)$
            	    \While {$|PQueue|+|\mathcal{A}_w| > h$}
            	        \State Remove the tuple $(\mathcal{E''},A'',Met_D'')$ from $PQueue$ with lowest $Met_D''$
            	    \EndWhile
            	\EndFor
            \EndWhile
		\EndFor
	\end{algorithmic}
\end{algorithm}

Algorithm~\ref{alg-1} selects a path for quantum state $\varrho$ between a quantum-user pair $S$ and $D$ with a specific width $w$. 
It is similar to Dijkstra's algorithm~\cite{dijkstra2022note}, but the goal is to maximize the entanglement rate instead of minimizing the length. 
$A$ is the path, and $Met_D$ is the entanglement rate of the final path. 
$S$ and $D$ are two quantum-users trying to share a state, where $S$ is a user as the source node and $D$ is the other user as the destination node. Each $Q_i\in\mathcal{Q}$ records the number of qubits in switch $v_i$. 
Starting from $S$, the algorithm finds an unvisited node with the largest entanglement rate. Then the algorithm enumerates the neighbor nodes to update the largest entanglement rate. After traversing all neighbor nodes, the algorithm marks this node as visited and repeats this process until $D$ is visited. 
The correctness of the algorithm relies on that the metric is monotonically decreasing, and we will skip the proof due to the space limitation. We can prove this by the contradiction method to show that adding one more node to the path will decrease the entanglement rate.    

To find multiple paths with the largest entanglement rates with different widths in a network, we propose Algorithm~\ref{alg-2} by combining Algorithm~\ref{alg-1} and the structure of Yen's algorithm. 
Yen's algorithm for finding k-shortest paths can employ any shortest path algorithm. 
The algorithm finds the shortest path, then makes deviation points for each hop to find the following potential shortest paths. Specifically, for a path $A=\{a_1,a_2,...,a_l\}$ includes $(a_i,a_{i+1})$ as a hop, the algorithm finds the shortest path $A'$ from $a_i$ to $a_l$, and make the new potential path by combining $\{a_1,a_2,...,a_i\}$ and $A'$. $A'$ cannot contain any nodes from $\{a_1,a_2,...,a_i\}$, nor can it contain the edge $(a_i,a_{i+1})$. 
There is a queue for storing the potential paths after generating potential paths. For each potential path, the length and the edges that cannot be used are also recorded, and potential paths generated by the current potential path cannot use these edges. The algorithm picks the potential path with the shortest length and repeats the above operations until all k-shortest paths are found or there is no more potential path. 

Algorithm~\ref{alg-2} can compute paths with the largest entanglement rates among all quantum-user pairs $\varrho_{SD}\in \mathcal{M}$. Yen's algorithm can employ the Algorithm~\ref{alg-1} as follows. 
The upper bound of the width, $\textsc{MAX\_WIDTH}$, is the maximum number of qubits in a single switch. 
Based on the `wider' principle, the algorithm starts to build paths from the largest width. 
Given a fixed width $w$, the algorithm finds $h$ paths with the largest entanglement rate and stores them in $\mathcal{A}_w$. The algorithm first finds a path with the largest entanglement rate. $PQueue$ is the queue to store potential paths with the highest entanglement rate. It stores tuples, and each of them includes three elements, an edge set, the path, and the entanglement rate. The algorithm finds the paths in $PQueue$ with the largest entanglement rate, and it is the next largest shortest path if it is not in the set $\mathcal{A}_w$. The algorithm then generates potential paths in the same way as Yen's algorithm. Finally, the algorithm stops when $A_w=h$ or there is no potential path in $PQueue$.

 

\subsubsection{Step II: Determine entanglement routes}
With a set of feasible paths from Step I, we propose an algorithm to select the set of routes which is denoted as $\mathcal{A}$ for the entanglement to maximize the network entanglement rate. The paths with the same shared quantum states between a quantum-user pair could be merged as a flow-like graph to save qubits in the network. The details are summarized in Algorithm~\ref{alg-3}. 

\begin{algorithm}
	\small
	\caption{Paths Merge Algorithm}
	\label{alg-3}
	\begin{algorithmic}[1]
		\Ensure $\mathcal{G}=(\mathcal{\overline{V}},\mathcal{E}), \mathcal{S}, \mathcal{D}, \mathcal{M}=\{\varrho_{SD}, S \in \mathcal{S}, D \in \mathcal{D}\}, \mathcal{Q},\newline \mathcal{A}_w,\forall w\in[1,\textsc{MAX\_WIDTH}]$
		\Require $\mathcal{A}$
		\ForAll {width $w$ from MAX\_WIDTH to 1}
		    \State Sort $\mathcal{A}_w$ by decreasing the order of metric
			\ForAll {$A\in \mathcal{A}_w$}
			    \State $good\xleftarrow{}1$
			    \ForAll {$i\in [0,l)$}
			        \If {Remaining qubits are not enough for edge $(a_i, a_{i+1})$ and no previous paths include edge $(a_i, a_{i+1})$ for $\varrho_{SD}$}
			        \State $good\xleftarrow{}0$
			        \EndIf
			    \EndFor
			    \If {$good=1$}
			        \State Record $A$ to $\mathcal{A}$
    			    \ForAll {$i\in [0,a_l)$}
    			        \If {No previous paths include edge $(a_i, a_{i+1})$ for corresponding pair $(a_0,a_l)$}
    			        \State Assign edge $(a_i, a_{i+1})$ width $w$ qubits to $\varrho_{SD}$
    			        \State Remove $w$ qubits from node $a_i$, $a_{i+1}$
    			        \EndIf
    			    \EndFor
    		    \EndIf
    		\EndFor
		\EndFor
	\end{algorithmic}
\end{algorithm}
We briefly explain Algorithm~\ref{alg-3} here. 
We first enumerate widths from high to low, then sort paths with the specific width in decreasing order of the entanglement rate. 
Paths connecting the same quantum state will be merged. 
We enumerate each edge of the path to check the remaining qubits at both endpoints of the edge (Line 3 to Line 9). 
For each path, there are two cases when merging it to previous paths. 
The first case is when the remaining qubits in both endpoints should be enough, in which case the path can be merged directly as the remaining qubits are enough. 
The second case is when there are some previous paths including some edges for the corresponding pair. 
In this case, the path is essentially a branch of previous paths, as 
there may exist flow-like graphs under $n$-fusion entanglement. The qubits in the previous edges can be shared with the current path, and the qubits will not be removed. 

\subsubsection{Step III: Utilize remaining qubits}
After running Algorithm~\ref{alg-3}, we have assigned most qubits in the network to build entanglement routes. 
However, there may be still a few qubits remaining in the network,
which can be assigned to improve the network entanglement rate. 
We will add the remaining qubits to selected routes from Algorithm~\ref{alg-3} to maximize the entanglement rate by increasing the width of quantum channels. 
The set of the newly added links is denoted as $\mathcal{A}'$. 

For a pair of qubits from two adjacent switches, a quantum link can be added to multiple routes from different shared quantum states. 
We will add the link to the route with the largest increase in entanglement rate compared to before. 
The procedure is outlined in Algorithm~\ref{alg-4}. 
Even though calculating the increased rate is with high time complexity, the total run time of the computation is acceptable because the number of remaining qubits after Algorithm~\ref{alg-3} in the graph is low. 


\begin{algorithm}
	\small
	\caption{Remaining Qubits Assignment Algorithm}
	\label{alg-4}
	\begin{algorithmic}[1]
		\Ensure $\mathcal{G}=(\mathcal{\overline{V}},\mathcal{E}), \mathcal{S}, \mathcal{D}, \mathcal{M}=\{\varrho_{SD}, S \in \mathcal{S}, D \in \mathcal{D}\}, \mathcal{Q}$
		\Require $\mathcal{A}'$
		\ForAll {$e_{ij}\in\mathcal{E}$}
		\While {$v_i,v_j$ have at least 1 qubit}
			\ForAll {$\varrho_{SD}$ in $\mathcal{M}$}
			\State Compute the increment of expected entanglement probability and record $\varrho_{SD}$ with the highest increment
			\EndFor
			\State Assign a 1 qubit of $v_i,v_j$ to $e_{ij}$ correspond to $\varrho_{SD}$, and record it in $\mathcal{A}'$
		\EndWhile
		\EndFor
	\end{algorithmic}
\end{algorithm}

The final entanglement routing algorithm is to combine the routes computed by Algorithm~\ref{alg-3} and the remaining qubits assignment computed by Algorithm~\ref{alg-4}. 
All routes employ the $n$-fusion entanglement method, which indicates that switches will entangle their interior qubits as many as possible to fuse links for one shared quantum state between a quantum-user pair when a route is given. 

\subsection{Time Complexity}
The complexity of Algorithm~\ref{alg-1} is $O(|\mathcal{\overline{V}}|\log |\mathcal{\overline{V}}|+|\mathcal{E}|)$. 
The time complexity of Algorithm~\ref{alg-3} is $O(\textsc{MAX\_WIDTH}*h|\mathcal{\overline{V}}|(|\mathcal{\overline{V}}|\log |\mathcal{\overline{V}}|+|\mathcal{E}|))$, which is composed by the number of loops for each width and Yen's algorithm. Yen's algorithm calls $h|\mathcal{\overline{V}}|$ times of Algorithm~\ref{alg-1}, where $h$ is the required number of shortest paths. The time complexity of Algorithm~\ref{alg-4} is $O(\textsc{MAX\_WIDTH}|\mathcal{E}|(|\mathcal{\overline{V}}|+|\mathcal{E}|)(|\mathcal{M}|+\textsc{MAX\_WIDTH})$.

The actual time cost is much lower for Algorithm~\ref{alg-4}, since most switches may not remain $\textsc{MAX\_WIDTH}$ qubits, thus the $\textsc{MAX\_WIDTH}|\mathcal{E}|$ term is loose. 

\section{Simulation Results} \label{sec: simulation}
In this section, we present the results of our simulations. We have implemented the proposed algorithms and compared their performance to existing methods. We have conducted extensive evaluations by varying multiple parameters to increase the reliability of the simulations. 

\subsection{Network Generation} \label{sec:simulation1}
To demonstrate differences in performance between our proposed algorithms and other existing works, we design controlled simulations under different network parameters. 

The default network settings are as follows. 
The different network parameters will be tested separately later. We generate the network through Waxman method~\cite{waxman1988routing}. 
The area of the quantum network is set as $10k \times 10k$ unit square, each unit may be considered as 1 kilometer~\cite{shi2020concurrent}.

Switches and quantum-users are nodes randomly placed in the area. 
The number of switches is set as 100.
The number of quantum states demanded shared by quantum-user pairs is set as 20. 
The edge generation follows the work~\cite{waxman1988routing}. 
Quantum-user nodes do not connect with other quantum-user nodes directly, and they are connected with switches directly. 
The distance of each edge is at least $\leq \frac{50}{\sqrt{|\mathcal{\overline{V}}|}}$, where $|\mathcal{\overline{V}}|$ is the number of nodes in the network.
The number of edges is determined by the average degree of switches which is set as 10.
We assume that each edge has enough capacity to serve quantum-users according to our assumption in the model. 
The main limitation of the network is switch capacity, i.e., the number of qubits in the solid memory, is set as 10. 
Considering the randomness of the network topology, 
We generate 5 random networks as a set and take the average value of the measured results, i.e., the entanglement rate. 
The successful entanglement swapping probability of the switch is set as 0.9 by default. 
The average successful entanglement probability over links follows  $P_{ij}=e^{-\alpha L_{ij}}$, where $\alpha$ is $10^{-4}$.

\subsection{Algorithm Benchmarks}
We compare the network performance with the following algorithms. 

\begin{itemize}
    \item \textsc{Alg-n-fusion}: We name our proposed entanglement routing algorithm as \textsc{Alg-n-fusion} that consists of four sub-algorithms. 
    \item \textsc{Q-Cast}: 
    This is a special version of \textsc{Alg-n-fusion} where $N=2$, which indicates that switches only take the classic swapping through BSMs. This version is very similar to Q-Cast proposed in ~\cite{zhang2021fragFmentation, shi2020concurrent}. Therefore, we name it \textsc{Q-Cast}.
    \item \textsc{Q-Cast-n}: 
    We apply Q-Cast from \cite{zhang2021fragFmentation, shi2020concurrent} to get paths. Then, we use Equation~\ref{eq: entanglement rate} to evaluate the network performance, assuming all paths take $n$-fusion. 
    \item \textsc{Baseline-1}(\textsc{B1}): We extend the algorithm in \cite{patil2021distance} from single pair to multiple pairs. For each pair, we run the algorithm once and remove the occupied resources. The result is the total entanglement rate of all pairs.
\end{itemize}

\subsection{Results}
\subsubsection{Main observations}
\ 
\noindent

\textbf{$n$-fusion versus the classic swapping.}
From simulations, we observe that given a fixed network with the same network resources, the performance of the network (i.e., entanglement rate) under $n$-fusion entanglement-swapping (i.e., GHZ measurements) significantly outperforms the case under the classic swapping method (i.e., BSMs). 
More specifically, \textsc{Alg-n-fusion}, \textsc{Q-Cast-n}  and \textsc{B1} can improve the network entanglement rate by up to 655\%, 198\% and 92\% compared with \textsc{Q-Cast}, respectively.  
This is because $n$-fusion is a more efficient swapping method that can better utilize network resources than classic swapping. 
Switches have the ability to fuse more quantum links, which can improve the success probability of entangling qubits of quantum-user pairs with the same network resources. 

\textbf{Performance under $n$-fusion.} 
Under the same network entanglement-swapping method with $n$-fusion in the same network, 
\textsc{Alg-n-fusion} can improve the network entanglement rate by up to 153\% and 293\% compared to \textsc{Q-Cast-n} and \textsc{B1} respectively. 
This indicates that our proposed \textsc{Alg-n-fusion} is the most efficient algorithm among them as it can fully utilize the network resources to improve the network performance. 
Moreover, \textsc{Alg-n-fusion} shows the incomparable performance when the link's successful entanglement probability and the switch's successful swapping probability are small, which is a case closer to reality. 
For example, $p=0.1$ from Figure~\ref{sim: p} and  $q=0.3$ from Figure~\ref{sim: q}. 
The reason for this is that \textsc{Alg-n-fusion} can efficiently utilize qubits through merging links to increase the entanglement rate. 


\subsubsection{Network generation methods} 
\ 
\begin{figure}[htbp]
\centerline{\includegraphics[width=0.27\textwidth]{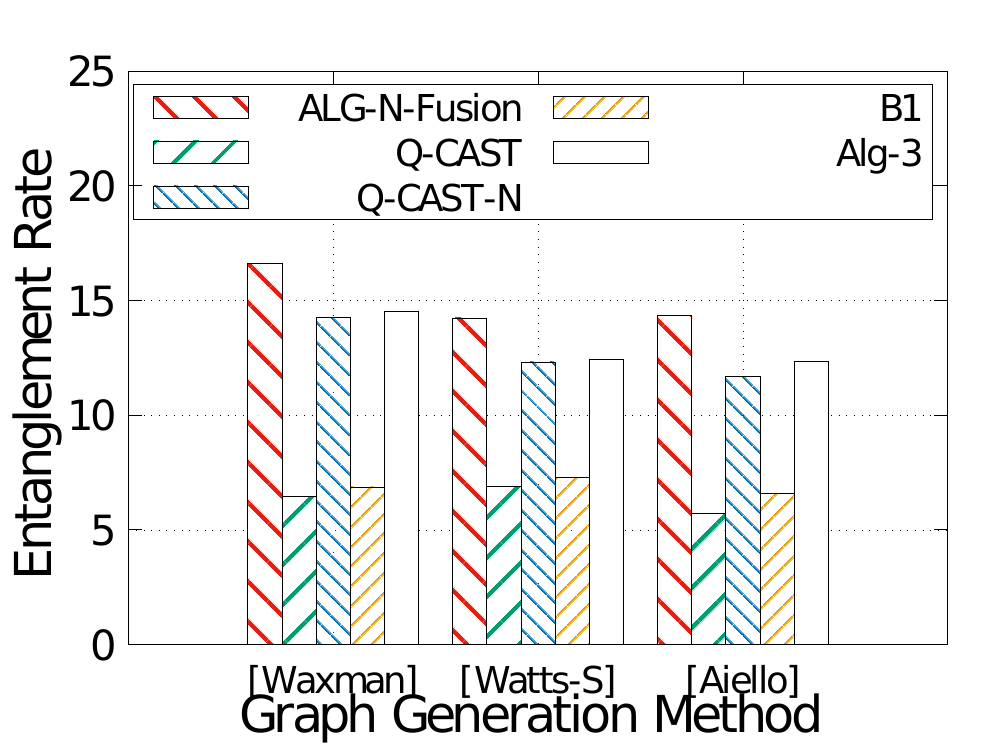}}
\caption{The network entanglement rate vs. different network generation methods. }
\label{sim: graph}
\vspace{-0.2in}
\end{figure}

We apply other two different methods to generate networks. The first one is the Watts-Strogatz method~\cite{watts1998collective}, which generates networks that exhibit properties similar to real-world communication networks.
The second one is Aiello method~\cite{volchenkov2002algorithm}, which generates scale-free power-law random graphs that resemble real-world network topologies.

Figure~\ref{sim: graph} presents the results of comparing network performance under three network generation methods. 
As shown, \textsc{Alg-n-fusion} 
achieves the highest network entanglement rate when compared to other benchmarks in networks generated by different methods. 
This suggests that our proposed algorithm can adapt to general network topologies and achieve good performance.

\subsubsection{Algorithm~\ref{alg-4} Performance} 
\ 

Figure~\ref{sim: graph} also shows the improvement brought by Algorithm~\ref{alg-4} by comparing the results between \textsc{Alg-n-fusion} and the network performance after only running Algorithm~\ref{alg-3} without Algorithm~\ref{alg-4}. 
It can be seen that Algorithm~\ref{alg-4} can improve the network entanglement rate up to 16.3\%. 
This is due to the fact that some qubits will not be selected for paths in Step I because of the width limitation. 
$n$-fusion can let these qubits be fully utilized for the entanglement in  Step III, thanks to Algorithm~\ref{alg-4}, leading to an improvement in network performance. 
We can conclude that Algorithm~\ref{alg-4} is an important component of our entanglement routing algorithm. 

\subsubsection{Quantum Parameters}
\

\begin{figure}[ht] 
\vspace{-0.1in}
    \centering
    \subfloat[]{
        \includegraphics[width=0.47\columnwidth]{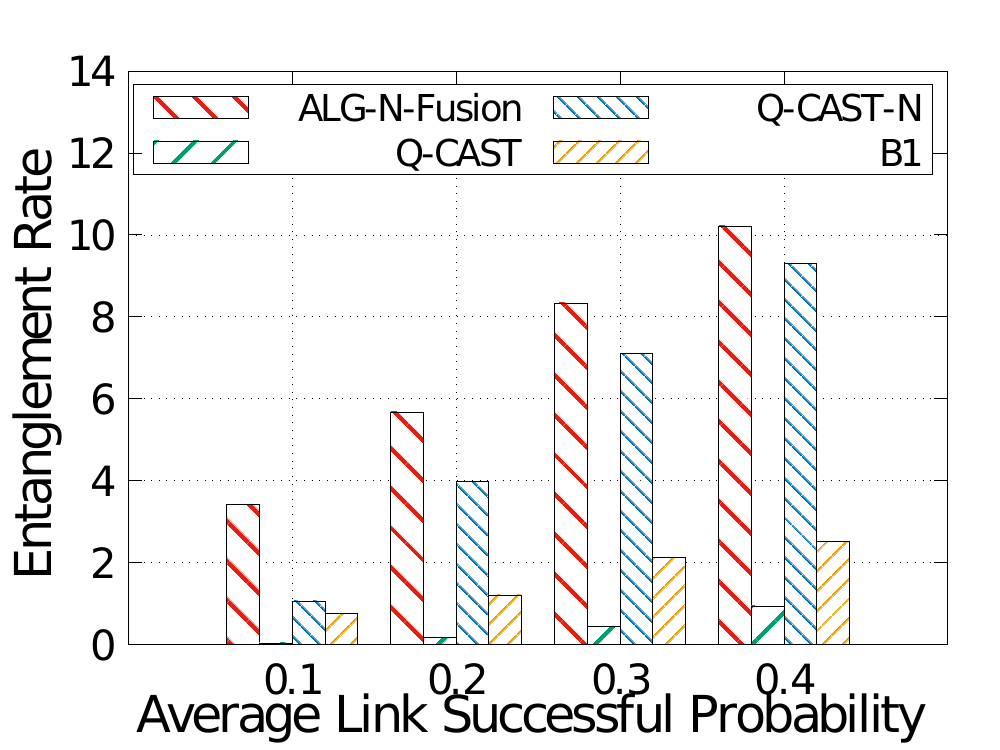} \label{sim: p}
    }
    \subfloat[]{
        \includegraphics[width=0.47\columnwidth]{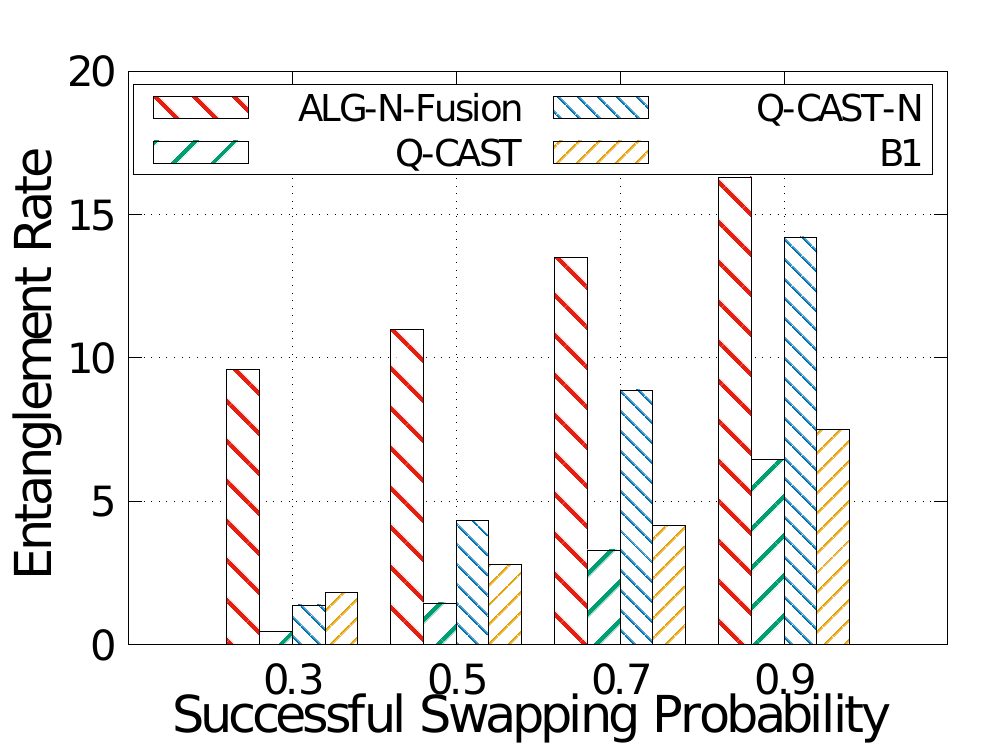}\label{sim: q}
    }
    \caption{(a)  The network entanglement rate vs. average quantum link successful entanglement probability (i.e., $p$).  (b) The network entanglement rate vs. switch entanglement-swapping probability (i.e., $q$).}
\end{figure}

Figure~\ref{sim: p} and Figure~\ref{sim: q} show the results of the network entanglement rates when varying the quantum link successful entanglement probability (i.e., $p$) and the entanglement-swapping probability (i.e., $q$) respectively. 

The results from Figure~\ref{sim: p} show that network entanglement rates by varying the average quantum link successful entanglement probability $p$. 
We assume that all links have the same $p$ to avoid the randomness brought by the network generation. 
When $p$ grows from 0.1 to 0.4, the network entanglement rate has an evident improvement by up to 775\%. 
\textsc{Alg-n-fusion} still outperforms other algorithms. 
The gaps among \textsc{Alg-n-fusion} and others is smaller when the $p$ is larger. This highlights the effectiveness of \textsc{Alg-n-fusion} in efficiently utilizing network resources to improve the network entanglement rate for the condition in reality where $p$ is small.   

As shown in Figure~\ref{sim: q}, the successful entanglement-swapping probability $q$ also has a significant impact on the network entanglement rate as the entanglement rate is closed related to $q$.   
An increase in $q$ leads to an increase in the entanglement rate.

\begin{figure}[ht] 
\vspace{-0.2in}
    \centering
    \subfloat[]{
        \includegraphics[width=0.47\columnwidth]{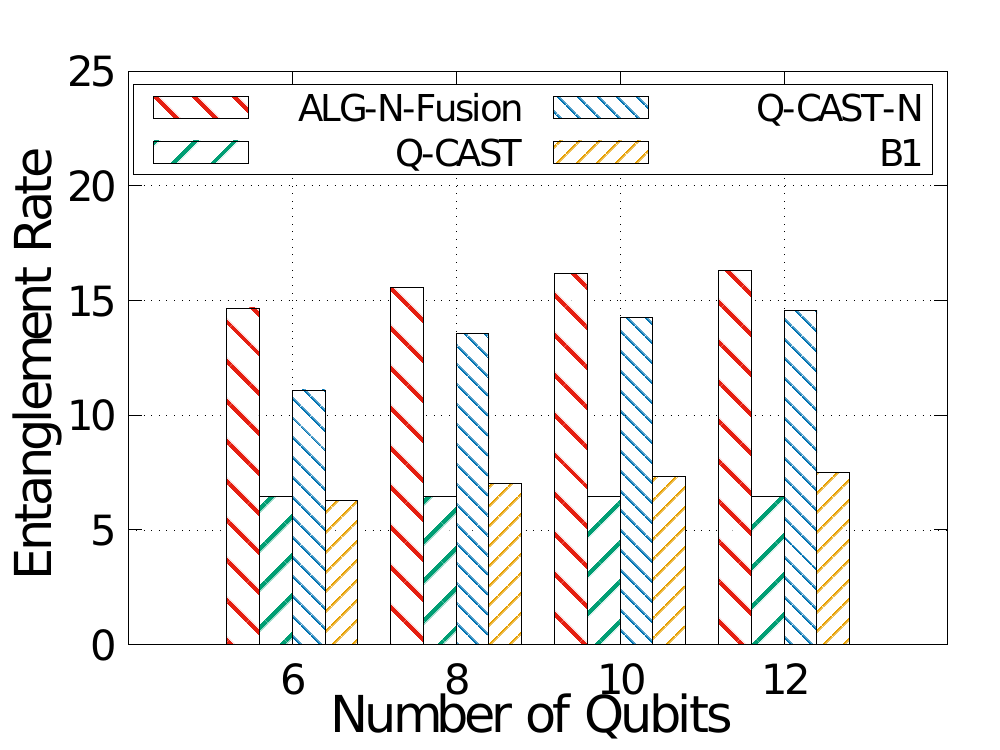} \label{sim: qubit}
    }
    \subfloat[]{
        \includegraphics[width=0.47\columnwidth]{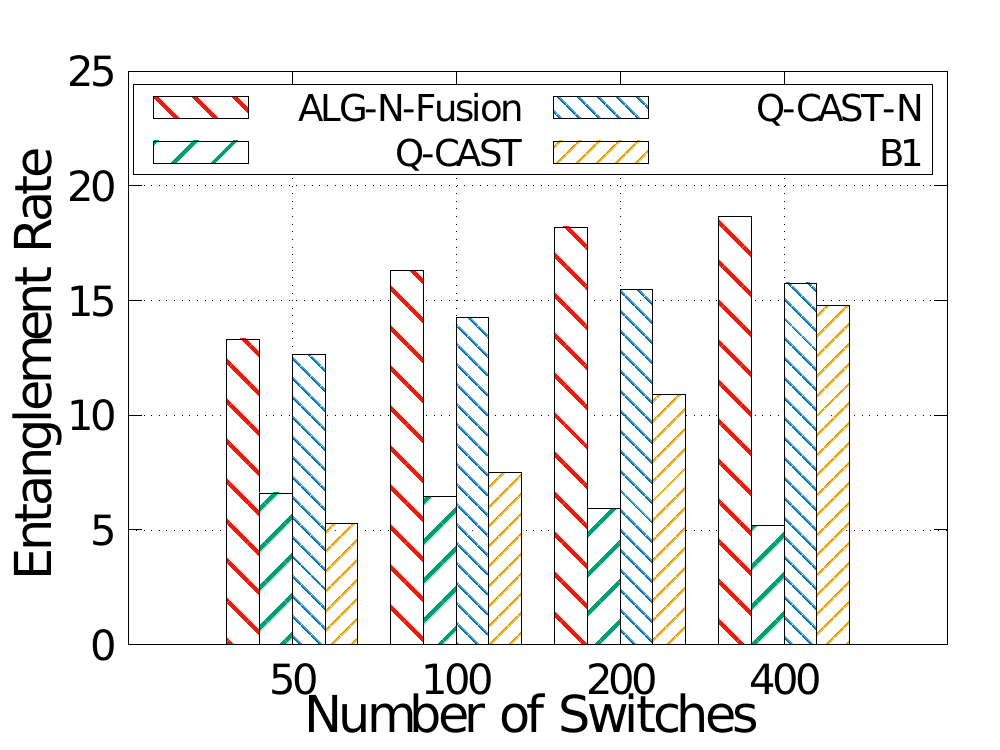}\label{sim: swtich}
    } \\
    \vspace{-0.1in}
     \subfloat[]{
        \includegraphics[width=0.47\columnwidth]{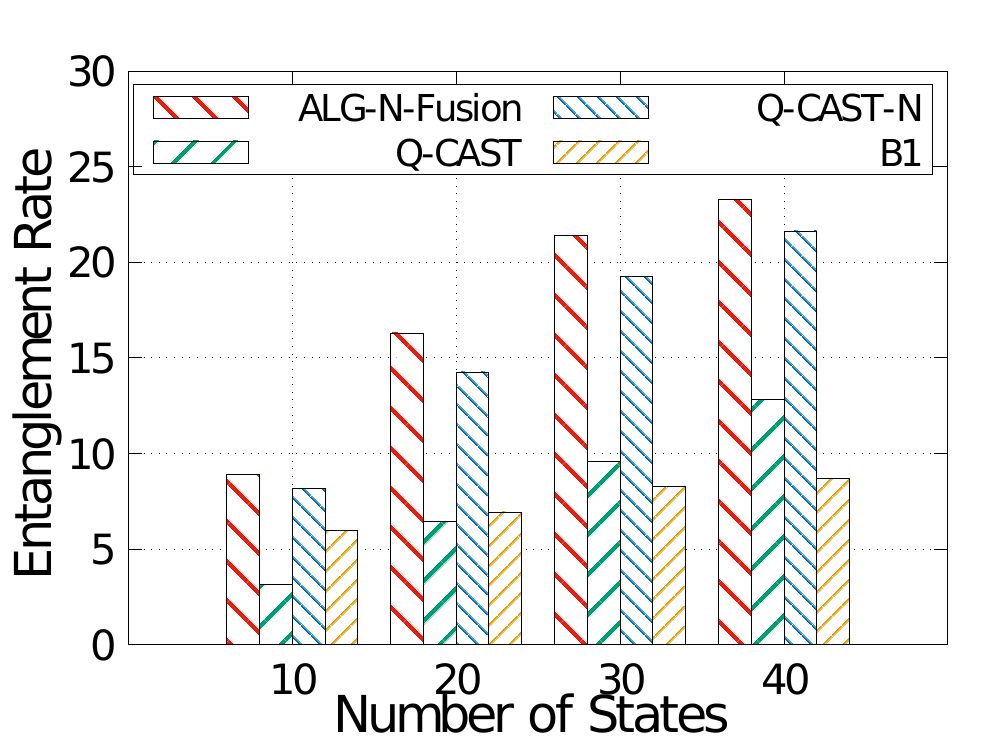} \label{sim: states}
    }
    \subfloat[]{
        \includegraphics[width=0.47\columnwidth]{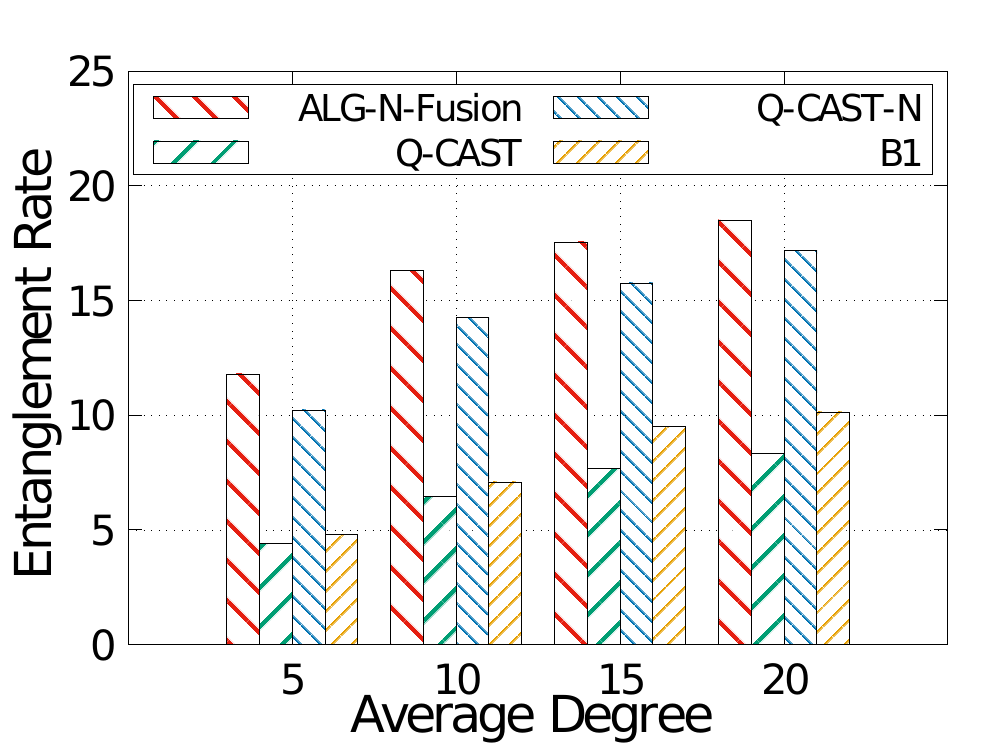}\label{sim: degree}
    } 
    \caption{(a)  The network entanglement rate vs. the number of qubits in a switch.  (b) The network entanglement rate vs. the number of switches. (c)  The network entanglement rate vs. the number of quantum states to be shared. (d)  The network entanglement rate vs. the average degree of a switch.}
\vspace{-0.1in}
\end{figure}

\subsubsection{Network Parameters}

We test four network parameters in our simulations: the number of qubits in a switch; the number of switches; the number of quantum states to be shared between quantum-user pairs; the average degree of a switch in the network. 
The results of varying these parameters are presented in Figure~\ref{sim: qubit},   Figure~\ref{sim: swtich},  Figure~\ref{sim: states}, and Figure~\ref{sim: degree} respectively. 

In this paper, we consider qubits in switches as the main limitation for entanglement routing. 
More switches and qubits in each switch can enlarge the total number of qubits in the network. 
In Figure~\ref{sim: qubit}, when the number of qubits in each switch is larger, the network performance is better.
In Figure~\ref{sim: swtich}, the entanglement rate increases obviously when the number of switches is larger. 
There is a special case. \textsc{Q-Cast} has fewer entanglement rates when the number of switches increases. This is because the routing metrics of \textsc{Q-Cast} suffer from larger distances and cannot efficiently utilize network resources.  
Figure~\ref{sim: states} shows that the entanglement rate increases with the number of quantum states to be shared because the demands from quantum-user pairs are increasing. 
Figure~\ref{sim: degree} plots the network entanglement rate with different average degrees of each switch. 
The network entanglement has an obvious increase when the average degree is larger. 
A larger number of the average degree in the network can add more connections between switches, which leads to more path selections for quantum users to share entanglement states. 
Results show that $n$-fusion can benefit a lot from this since more edges give $n$-fusion more flexibility to fuse links. 
Additionally, we conclude that our proposed algorithm can fit well with general network topologies and can efficiently utilize network resources.

\section{Related Works} \label{sec: related work}

\textbf{System implementation:}
Some research labs and companies have constructed several trial quantum networks for the quantum key distribution or real qubit transmission such as DARPA quantum system~\cite{elliott2005current}, SECOQC Vienna QKD system~\cite{peev2009secoqc}, Tokyo QKD system~\cite{sasaki2011field},
the mobile quantum system~\cite{liu2021optical}, and the integrated satellites~\cite{chen2021integrated}. 
Due to the hardware limitation, there are still no large-scale quantum networks that can be widely applied in reality.  

Currently, existing studies mainly focused on the theoretical aspect of quantum networks, and the main evaluation methods were based on numerical evaluation or simulation based on virtual simulators.    

\textbf{Entanglement routing under BSMs:}
This group of studies considered the classic swapping method based on BSMs in quantum networks. 
Vardoyan \textit{et al.}~\cite{vardoyan2019stochastic} studied the theoretical performance of processor capacity and memory occupancy distribution for a single processor with multiple quantum-users. 
Shchhukin \textit{et al.}~\cite{shchukin2019waiting} analyzed the average waiting time for a single entanglement path based on Markov chain theory. 
Pant \textit{et al.}~\cite{pant2019routing} proposed a local routing policy for independent processors both in single-flow and multi-flow. 
Das \textit{et al.}~\cite{das2018robust} presented a routing algorithm for two sets of quantum-users in a Bravais lattice topology. 
Li \textit{et al.}~\cite{li2021effective} studied the flow-based system performance in a lattice network. 
\cite{chakraborty2019distributed} proposed a greedy routing design in ring and grid networks. 
These works considered routing design in quantum computing systems with special topologies. 

Shi \textit{et al.}~\cite{shi2020concurrent,zhang2021fragFmentation} proposed routing algorithms for a random graph to maximize network throughput.  The proposed algorithms were greedy-based that selected the path with the largest throughput until no feasible paths. 
\cite{chakraborty2020entanglement, qiao2022quantum} considered the fidelity as an entanglement constraint. 
Zeng \textit{et al.}~\cite{zeng2022multi} designed routing algorithms to maximize the number of served quantum-user pairs and the expected throughput simultaneously. 

\textbf{Entanglement routing under $n$-fusion:}
Patil \textit{et al.}~\cite{patil2021distance} found an interesting conclusion that in a grid/lattice network where switches took $n$-fusion entanglement-swapping, the entanglement rate between one pair of quantum-users is not dependent with the distance. 
Later, they extended this finding to a model with a space-time multiplexed method~\cite{patil2022entanglement}.   
However, this conclusion was based on the percolation theorem~\cite{essam1980percolation} where the graph had a special structure. They only discussed a condition with one pair of quantum users.

\section{Conclusion}\label{sec: conclusion}
In this paper, we have proposed a general entanglement routing model of quantum networks with general topologies where switches take $n$-fusion, a general entanglement-swapping method. We have proposed efficient algorithms for multiple quantum-user pairs to maximize the network entanglement rate. 
Extensive numerical evaluations have shown that our proposed algorithms enable the network under $n$-fusion to have a better performance compared with existing works. 
This paper lays the groundwork for an in-depth exploration of $n$-fusion as a method for enabling quantum applications through swapping. It is our aspiration that this work sparks interest and drives forward-thinking research in this field.

\section*{Acknowledgment}

This work is supported in part by US National Science Foundation under grant numbers 1717731, 1730291, 2231040, 2230620, 2214980, 2046444, 2106027, and 2146909.

\bibliographystyle{IEEEtran}
\bibliography{reference}


\end{document}